\documentclass[preprint,12pt]{elsarticle}

\usepackage{lipsum}
\makeatletter
\def\ps@pprintTitle{%
	\let\@oddhead\@empty
	\let\@evenhead\@empty
	\def\@oddfoot{}%
	\let\@evenfoot\@oddfoot}
\makeatother

\usepackage{subfiles} 

\usepackage{graphicx}
\usepackage{lscape} 

\usepackage{amssymb}

\usepackage{hyperref}
\usepackage{doi}

\usepackage{lineno}
\usepackage{mhchem}
\usepackage[english]{babel}
\usepackage{natbib}
\biboptions{numbers, sort&compress}
\emergencystretch=\maxdimen
\addto\captionsenglish{} 

\usepackage[section]{placeins} 
\usepackage{lscape}  

\usepackage{listings}
\usepackage{graphicx} 
\usepackage[table,xcdraw]{xcolor}
\usepackage{multirow}
\usepackage{longtable} 

\begin{document}

\begin{frontmatter}


\title{Dynamic Chemical Model for \ce{H2}/\ce{O2} Combustion Developed Through a Community Workflow}



\author[UCB]{James Oreluk}
\address[UCB]{\mbox{Department of Mechanical Engineering, University of California, Berkeley, CA 94720, USA}}
\author[NCSTATE]{Craig D. Needham}
\address[NCSTATE]{Department of Chemical \& Biomolecular Engineering, North Carolina State University, Raleigh,
	NC 27695-7905, USA}
\author[KAUST]{Sathya Baskaran}
\author[KAUST]{S. Mani Sarathy}
\address[KAUST]{King Abdullah University of Science and Technology, Thuwal, Saudi Arabia}
\author[COLUMBIA]{Michael P. Burke}
\address[COLUMBIA]{Department of Mechanical Engineering, Department of Chemical Engineering, and Data Science
	Institute, Columbia University, New York, NY 10027, USA}
\author[NEU]{Richard H. West}
\address[NEU]{\mbox{Department of Chemical Engineering, Northeastern University, Boston, MA 02115, USA}}
\author[UCB]{Michael Frenklach}
\author[NCSTATE]{\mbox{Phillip R. Westmoreland}}

\begin{abstract}
Elementary-reaction models for \ce{H2}/\ce{O2} combustion were evaluated and optimized through a collaborative workflow, establishing accuracy and characterizing uncertainties. Quantitative findings were the optimized model, the importance of \mbox{\ce{H2 +}\mbox{\ce{O2}$(1\Delta)$}\ce{= H + HO2} } in high-pressure flames, and the inconsistency of certain low-temperature shock-tube data. The workflow described here is proposed to be even more important because the approach and publicly available cyberinfrastructure allows future community development of evolving improvements. The workflow steps applied here were to develop an initial reaction set using Burke et al. [2012], Burke et al. [2013], Sellev\aa g et al. [2009], and Konnov [2015]; test it for thermodynamic and kinetics consistency and plausibility against other sets in the literature; assign estimated uncertainties where not stated in the sources; select key data targets (“Quantities of Interest” or QOIs) from shock-tube and flame experimental data; perform conventional sensitivity analyses of QOIs with respect to Arrhenius pre-exponential factors; develop surrogate models for the model-predicted QOI values; evaluate model-vs.-data consistency using Bound-to-Bound Data Collaboration; and optimize model parameters within their estimated uncertainty bounds (feasible set). Necessary data and software for such analyses were developed and are publicly available through the PrIMe cyberinfrastructure. This community workflow proved to be a means to reveal inconsistencies, improvements, and uncertainty bounds. Even more significantly, it is a means of revealing which parameters and experimental findings are inconsistent with the larger body of work from the community and, thus, of designing new experiments and parameter calculations.
\end{abstract}


\end{frontmatter}


\section{Introduction}
\label{intro}

Modeling combustion has proven to be a powerful tool for understanding combustion chemistry and for designing combustor improvements. In general, a chemical reaction set with species thermochemistry and kinetics (a ``mechanism''), transport properties, and a physical model of the combustion device are required. Making advances increasingly requires capturing the combined uncertainty of model parameters, model structure, and experimental data, addressing the questions: Do the data and model agree? What are the uncertainties? Which are the best parts of the model and/or the experiment to re-assess?

Here, the kinetics of \ce{H2}/\ce{O2} combustion has been used as an important scientific and technological subject and as a means of developing and testing the collaborative framework. Given its role as a classic, apparently simple, combustion kinetic model and as a core foundational component to any hierarchically developed kinetic model for all hydrocarbon and oxygenated fuels, the \ce{H2}/\ce{O2} model has historically been the topic of extensive attention, including a number of studies even within the last five years \cite{hong2011improved, you2012process, burke2012comprehensive, keromnes2013experimental, varga2015optimization, konnov2015role}. Present understanding, both quantitatively and qualitatively, continues to evolve as new theoretical, experimental, and modeling studies \cite{hong2011improved, you2012process, burke2012comprehensive, keromnes2013experimental, varga2015optimization, konnov2015role, ellingson2007reactions, burke2010negative, turanyi2012determination, burke2013quantitative, verdicchio2015highlevel} shed further light into the \ce{H2}/\ce{O2} model. Such high levels of sustained attention and continual discovery/refinement of the model itself emphasize the need for a dynamic approach to kinetic model development that readily incorporates new information. Quantification of remaining uncertainties and identification of the largest sources of remaining uncertainties are likewise subject to uncertainty, such that different researchers might justifiably assign different uncertainties to the same data, thus guiding their own distinct path toward improving the model, quantifying remaining uncertainties, and/or reconciling initial inconsistencies within a model-data framework.

In this regard, an uncertainty-quantification (UQ) approach that is both accessible by the entire scientific community and adaptive appears particular worthwhile. Here, we demonstrate such an approach, applied to the \ce{H2}/\ce{O2} question. Analogous to an adaptive simulation where a coarse initial grid is employed and then subsequently refined adaptively at locations found to be most important, an adaptive approach to UQ employs simple, relatively crude treatments of model uncertainty in the initial step. Thereafter, the initial analyses are used to prioritize further refinements in the treatments of model uncertainty. In the present work rate-constant pre-exponential factors are analyzed as the uncertain model parameters, and reported and approximate uncertainty estimates are applied for the experimental observables.

Implementation of the first stage of this adaptive approach applied to \ce{H2}/\ce{O2} chemistry, described below, has already yielded promising indications that this UQ framework can identify both well-recognized model/data inconsistencies (e.g., modeling low-temperature shock-tube ignition using homogeneous, constant-volume models) and suggest yet-unrecognized scientific interpretations and remaining structural uncertainties (e.g., an unexpected role of \mbox{$\text{O}_2(\text{a}^1\Delta\text{g})$} pathways in high-pressure flames).

\section{Theory and Procedures}
\subsection{Community workflow}

\begin{figure}[!htb]
	\centering\includegraphics[scale=0.35]{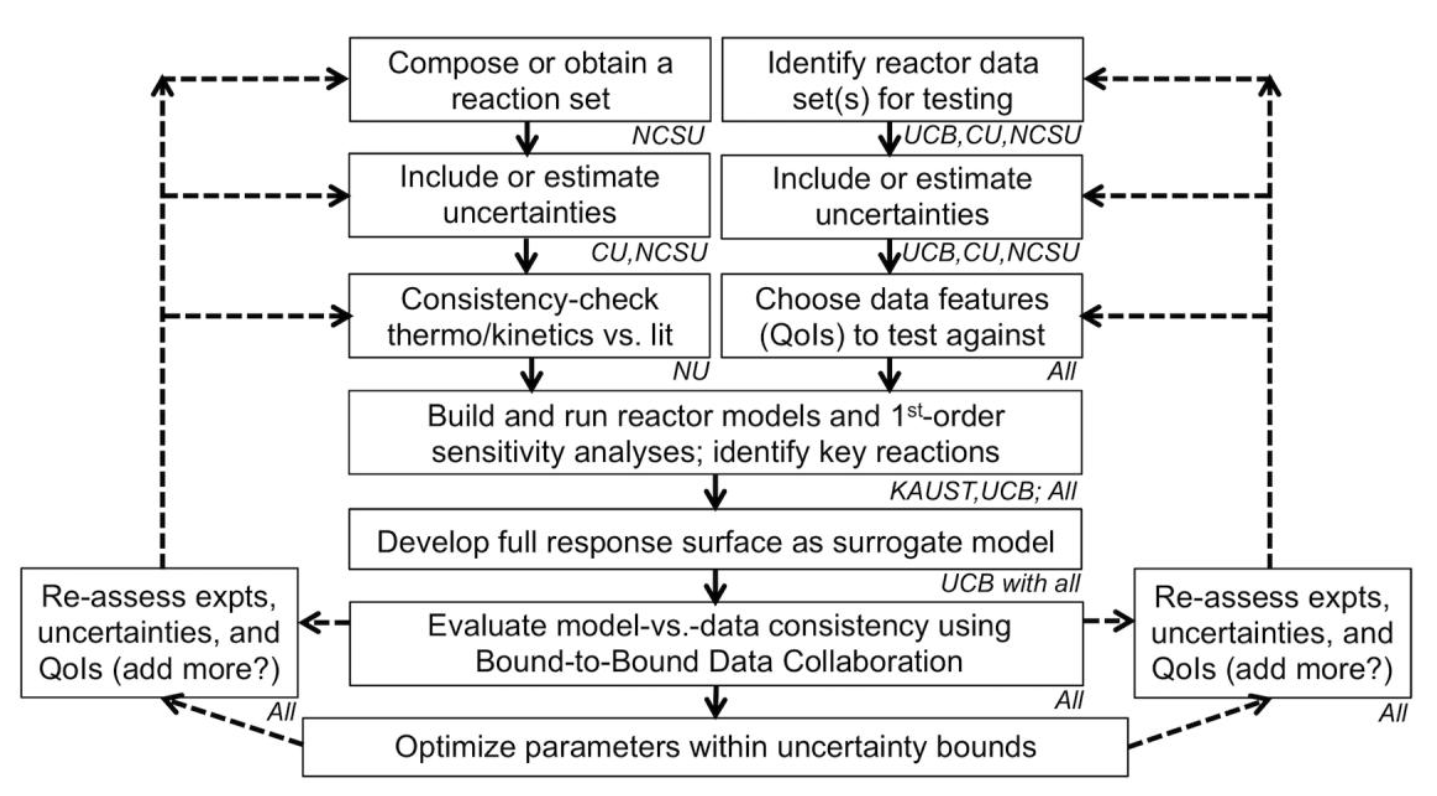}
	\caption{Workflow for initial and dynamic community development of models (with area leads)}.
	\label{fig:workflow}
\end{figure}

The community workflow summarized in Figure \ref{fig:workflow} is applicable to mechanism development with UQ generally. Leading roles in the present activity are shown to illustrate how this workflow implementation enables both individual autonomy and team (or group) collaboration. 

Beginning with the objective of addressing \ce{H2}/\ce{O2} kinetics, parallel initial tasks were to choose a base reaction set with thermochemistry and rate coefficients while archiving as many reactor-data sets as possible; both tasks required identifying or estimating parametric uncertainties. For checking reaction-set parameters, new codes were used to scan a wide range of literature values. Influential reactions were identified, and surrogate models were built for every Quantity of Interest (QOI: a characteristic feature or an attribute of selected experimental data to be tested). The consistency of the model (with uncertainties) vs. experimental QOIs (with uncertainties), along with sensitivity to the uncertainty bounds, was determined using Bound-to-Bound Data Collaboration (B2BDC) \cite{frenklach2004collaborative, feeley2004consistency, seiler2006numerical, russi2008sensitivity, russi2010uncertainty}. To attain consistency, the QOI bounds needed to be reassessed; after that, the model parameters were optimized. Aspects of this workflow were used previously in developing the GRI-Mech methane-combustion models in the 1990s \cite{smith1999gri}, but the advent of computational and data-science tools for treating uncertainty and larger data sets allows significant advances beyond GRI-Mech.

\subsection{Data archiving}

Kinetic model data (thermo, transport, rate constants) and experimental data (shock tubes and flames) were archived in the PrIMe Data Warehouse, an online XML database part of the PrIMe cyberinfrastructure \cite{frenklach2007transforming, PrIMe}. The PrIMe Data Model provides structure to data to be archived and creates relationships among data records, where each record is assigned a unique identifier (PrIMe ID). This allows data of published values, reported uncertainties, and metadata such as equipment diagrams to be interlinked to their respective bibliography references.

\subsection{Parameter consistency vs. literature}

The thermochemistry and kinetics data were subjected to testing for internal and literature consistency. When merging and updating kinetic models, it is easy to introduce errors, for example in the units of a rate coefficient. To detect such errors, and to inform the estimation of parameter uncertainties, the species thermochemistry and reaction rate expressions were compared against fifty kinetic models of combustion from various websites (AramcoMech, USC-Mech, GRI-Mech, LLNL) and published literature, including nineteen publications from Combustion and Flame (2013-2015) and all supplementary material of the 2013 Proceedings of the Combustion Institute. Using a recently developed tool \cite{lambert2015identification}, the names of the species representing the H/O chemistry were identified in each model so those equivalent reactions could be found in different models. The species enthalpies and reaction rate coefficients were then evaluated at a range of temperatures and pressures and compared. 

\subsection{QOI selection}
QOIs are carefully selected experimental observables used to characterize the phenomena one would like to model. Examples of QOIs include ignition delay, the peak value of a species profile, laminar flame speed, and time/location of a peak value. In the present study, QOIs are stored in the PrIMe Data Warehouse as ``Data Attributes'' files, which fully describe the experimental feature, conditions, uncertainty bounds, and transformations required to obtain the QOI value.

\subsection{QOI evaluation}
Simulations of shock-tube QOIs were performed with the PrIMe internal PFR component, as described in You et al. \cite{you2012process}.

Simulations of flame QOIs were performed using a cloud-based tool, CloudFlame \cite{reyno2015cloudflame, goteng2014hybrid}, coupled with the PrIMe Workflow Application (PWA) cyberinfrastructure. Briefly, the inputs for the QOI simulation, such as the chemical kinetic model, thermodynamic data, transport data, experimental conditions (temperature, pressure, mixture composition), and solver parameters (grid points, tolerances), are stored in a PrIMe Interface file (HDF5) and passed to CloudFlame via a web-service application. The HDF5 file also prescribes the output required for the simulations (e.g., laminar flame speed, OH concentration, mass burning rate).

Upon receipt, CloudFlame executes a Python script that reads the HDF5 file and launches a simulation for the prescribed conditions using the open-source software package CANTERA \cite{Cantera}. After completing the simulation, another Python script encodes the desired output(s) into an output HDF5 file and returns it to the PWA via a web-service application. For sensitivity analysis, the building of surrogate models, and optimization, PWA sends CloudFlame a single HDF5 with a matrix of kinetic parameters together with the simulation conditions for each, and upon completion of the simulations, a single HDF5 file with the results is returned.

Prior to activating the remote flame-simulation module, we verified that the CANTERA simulations provided results within 2\% of analogous simulations performed using CHEMKIN-PRO \cite{designchemkin} (comparisons available in Supplementary Material). Simulations included multicomponent transport and thermal diffusion (Soret effect). Gradient and curvature criteria were tight enough to provide a solution with no fewer than 1500 grid points.

Approximately 4700 premixed-laminar-flame-speed simulations were conducted using this workflow module. Simulations were performed in parallel using 64 compute nodes on a high-performance computing cluster at KAUST.

\subsection{Bound-to-Bound Data Collaboration}
The analysis is based on the B2BDC framework of uncertainty quantification (UQ) \cite{frenklach2004collaborative, feeley2004consistency, seiler2006numerical, russi2008sensitivity, russi2010uncertainty, frenklach2007transforming}, an optimization-based framework for combining models and experimental data from multiple sources to explore their collective information content. This framework calls for an access to all available data, advocated and demonstrated here through group activity.

The B2BDC framework identifies a set of parameter combinations (the ``feasible set'') that guarantees the model predictions will be within the ranges of uncertainties of all pertinent experimental observations. This framework is based on rigorous mathematics \cite{seiler2006numerical}, designed for integrating multiple and even disparate sets of experimental data to validate and discriminate models quantitatively and systematically.

The system in question is represented by a numerical model, $M(x)$, with parametric dependence on a set $x$ of $N$ given parameters (here, rate coefficients) that have quantified uncertainties. Thus, each $x_i$ is constrained to a one-dimensional interval $[x_{i,\text{min}}, x_{i,\text{max}}]$, and the full set of bounding values creates an $N$-dimensional hypercube $\mathcal{H}$. Note that the model $M(x)$ is not restricted to the chemical mechanism. It can incorporate any physical parameters of the reactor model and even the ``Instrumental Model'' parameters for converting raw data, such as calibration factors \cite{yeates2015integrated}. Experimental data are represented by the set of QOIs. Each $e$-th measurement has a reported value QOI$_e$ and lower and upper uncertainty bounds, $[L_e, U_e]$.

To be considered in agreement, the model $M(x)$ must produce outputs that are ``consistent'' with the experimental QOI set; i.e., they must be within the experimentally observed bounds. Thus, the feasible set of parameters $\mathcal{F}$ is a representation of the complete collaborative information contained in a data-model system, and questions in the B2BDC framework are posed as optimization problems over the feasible set. This reliance raises the question of dataset consistency: is $\mathcal{F}$ non-empty? To assess it numerically, a consistency measure was introduced \cite{feeley2004consistency} that quantifies how much the constraints can be tightened while still ensuring the existence of a set of parameter values whose associated model predictions match the experimental QOIs within the bounds. The data-model system is consistent if the consistency measure is nonnegative, and it is inconsistent otherwise. 

An integral part of the present B2BDC framework, which makes the approach practical, is the approximation of the detailed model output $M_e(x)$ for given QOI$_e$ by a quadratic surrogate model. Although doing so introduces some surrogate fitting error, the errors are controlled to be sufficiently small and are accounted for by increasing the experimental bounds, $L_e$ and $U_e$. When the model is fully consistent, its parameters can be optimized by using a least-squares minimization or more sophisticated objective function, such as the LS-F method \cite{you2012process} that was employed in the present study. It is a least-squares minimization constrained to the feasible set, which ensures that the best-fit vector $x$ predicts QOI values within their respective uncertainty bounds. 

\section{Results}
\subsection{Choice of QOIs}
In the present study, ignition delay times were used as QOIs for shock-tube experiments and laminar flame speeds as QOIs for laminar-flame experiments. For a shock-tube experimental series, three QOIs were taken from the log($\tau_{\text{ign}}$)-vs.-1/T fit of the experimental data at quarter, half, and three-quarter temperature. The total collection was 124 QOIs spanning 12 shock-tube studies (114 QOIs) and two flame studies (10 QOIs). 

Initial uncertainties assigned to the targets were based on the source authors, our estimates, and/or statistical analyses of the data. Reassessment of these initial uncertainties takes place when model-vs.-data consistency is sensitive to them. Later iterations of the present approach will benefit from the inclusion of a larger number of experimental measurements of a wider array of experimental observables.

\subsection{Base and modified mechanisms}
To test different aspects of the \ce{H2}/\ce{O2} kinetics, three models were created from literature sources. The base mechanism (Mechanism1) was taken from Burke et al. \cite{burke2012comprehensive}. This mechanism was chosen based on its high quality of performance when compared to both shock-tube and flame data. Each reaction rate from the mechanism was traced to the publication where the rate coefficient was measured or calculated and entered into the PrIMe database in its original form. The intent was to preserve the significant figures given in the original publication, record cited uncertainties accurately, and avoid any unit-conversion errors in subsequent publications. To match the reaction direction in the PrIMe-archived kinetics database, rate coefficients were reversed as necessary using the NASA-polynomial thermochemistry given in \cite{burke2012comprehensive}. To allow for complex temperature dependencies, the reverse rates were evaluated at 50 temperatures between 294 and 2000K and three-parameter modified-Arrhenius expressions fitted by least-squares regression. For third-body and falloff reactions, both low- and high-pressure limits were reversed, and Troe or Lindemann parameters remained unchanged. This analysis was performed using methods in the RMG-Py software \cite{generator2016automatic}. The resulting PrIMe-archived Mechanism1 \cite{orelukMech1} was used principally to ensure that the results shown in the original publication could be replicated with the PrIMe-workflow simulations. 

Mechanism2 \cite{orelukMech2} consisted of Mechanism1 with parameter updates from Burke et al. \cite{burke2013quantitative} and Sellevåg et al. \cite{sellevaag2009kinetics}.

Mechanism3 \cite{orelukMech3} added kinetics for ozone, \ce{O(^1D)}, \mbox{$\text{O}_2(\text{a}^1\Delta\text{g})$}, and \ce{OH^*} from Konnov \cite{konnov2015role}. Mechanism2 was used mainly for comparisons with the more-complete Mechanism3, differentiating the contributions of the excited species and the kinetics updates to the base. In all cases, the reaction rate constants employed in Mechanism3 are directly adopted from rate assessments and/or fundamental studies and either lie within literature, reviewed uncertainties (\ce{O + H2 = OH + H}) (e.g., \cite{baulch2005evaluated}) or are based on more recent studies \cite{burke2013quantitative, sellevaag2009kinetics, fernandez2002vtst, srinivasan2006thermal}.

Using the PrIMe B2BDC tool, described above, Mechanism3 was analyzed and optimized to create an optimized Mechanism4, designated as DynamicMech151203 \cite{orelukDynamicMech} to emphasize that it is not a version, but a dynamically improvable set. All mechanisms are provided in the Supplementary Material as CANTERA text-input format.

\subsection{Parametric consistency with the literature}
The thermochemistry of H/O chemistry proved quite consistent across the $\sim$50 models inspected, with the largest ranges in $\Delta_f$H$^\circ$ being for excited \ce{OH^*} (14 kJ/mol), \ce{HO2} (5 kJ/mol) and \ce{OH} (2.5 kJ/mol) radicals. Mechanism3 had $\Delta_f$H$^\circ$ values within 0.25 kJ/mol of the median for everything except \ce{OH^*}, which was between the only two other values found.

For the kinetics-consistency checks, very few significant deviations or errors were detected in the literature models for O/H chemistry because it is generally a sensitive and well-studied submechanism. Reactions were not reversed, so reactions described in opposite directions in different models were not compared. Discrepancies are generally larger at lower temperatures and lower pressures. At 500 K and 0.01 bar, for example, of the 21 reactions in ten or more models including Mechanism3, nine have standard deviations of at least log$_{10}(k)=1.0$ and a range of at least 7.0, while at 1500 K and 100 bar, all but one reaction have a standard deviation below log$_{10}(k)=0.5$ and a range below 2.0. Table \ref{table:rxnspeed} summarizes the cases where the rate differs from the median of the literature values at 1000 K and 0.01 bar for Mechanism3 and at least three other models.

\begin{table}[!htb]
	\centering
	\caption{Reactions occurring in more than three models that differ from median literature values by
		more than 10\% at 1000K, 0.01 bar}
	   \bigskip
	\label{table:rxnspeed}
	\resizebox{\textwidth}{!}{%
		\begin{tabular}{lcc}
		\multicolumn{1}{c}{\textbf{Reactions}}           & \textbf{\# of models} & \textbf{Difference from median} \\ \hline
			$\ce{OH + OH = H2O2}$        & 16                    & 2.93x slower                    \\
			$\ce{HO2 + OH = H2O + O2}$   & 48                    & 2.17x slower                    \\
			$\ce{HO2 + HO2 = H2O2 + O2}$ & 40                    & 1.81x slower                    \\
			$\ce{O + H2 = H + OH}$       & 49                    & 1.73x slower                    \\
			$\ce{OH + OH = O + H2O}$     & 27                    & 1.33x slower                    \\
			$\ce{HO2 + O = O2 + OH}$     & 46                    & 1.26x faster                    \\
			$\ce{H2O = H + OH}$          & 15                    & 1.11x slower                   
		\end{tabular}%
	}
\end{table}

\subsection{B2BDC analysis}
Each QOI was first analyzed for self-consistency before undertaking mutual consistency with all experiments; a QOI is considered self-inconsistent when no point $x$ in $\mathcal{F}$ can produce a QOI value that falls within that QOI's uncertainty bounds. Thirty-six shock-tube QOIs were found to be self-inconsistent. Experimental uncertainty bounds of all shock-tube QOIs were then re-estimated as twice the standard deviation from the fitted experimental data. This method represented the variance seen in the experimental data, but it caused ten QOIs to have implausibly tight bounds (below 5\% uncertainty). Relaxing the experimental bounds of the self-inconsistent targets to a conservative relative upper and lower bound of a factor of 2 brought ten QOIs to self-consistency. The remaining self-inconsistent QOIs were removed from the analysis, leaving them for future investigation.

\begin{figure}[!htb]
	\centering\includegraphics[height=0.50\linewidth]{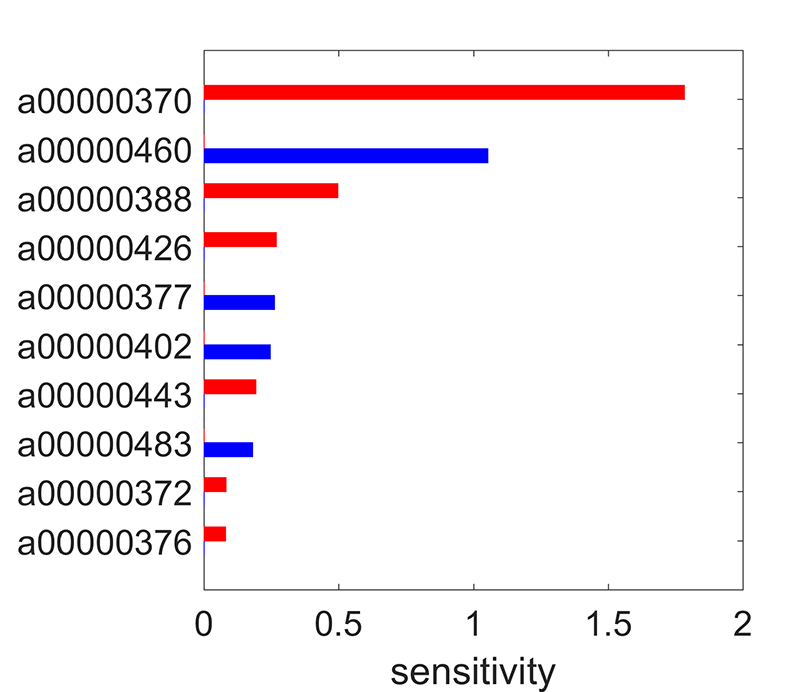}
	\caption{Normalized sensitivity of the consistency measure with respect to QOI lower (black/red) and upper (gray/blue) bounds. The ten most sensitive QOIs are all shock-tube data except a00000483, which is a flame speed experiment. QOI codes and conditions are detailed in Supplementary Material.}.
	\label{fig:sens}
\end{figure}

The system of the surviving self-consistent shock-tube QOIs and the model was still found to be inconsistent. To resolve this inconsistency, we examined the sensitivity of the consistency measure to each of the QOIs and parameter bound. High sensitivity was obtained for the experimental uncertainty bounds of shock-tube experiments, as shown in Figure \ref{fig:sens}. A relative change $>5\%$ was required to the upper or lower bound of 25 shock-tube QOIs to resolve the inconsistency. These QOIs were also withheld for future opportunity to re-evaluate their uncertainty bounds, surrogate approximations, and model-form uncertainty. The other QOIs were included in the model optimization.

\section{Discussion}
Two outcomes illustrate both the ability of the consistency analysis to identify data inconsistencies and the ability of the overall approach to identify new scientific interpretations and highlight yet unrecognized structural and parametric uncertainties.

\subsection{A role of $\text{O}_2(\text{a}^1\Delta\text{g})$ chemistry}
\begin{figure}[!htb]
	\centering\includegraphics[height=0.65\linewidth]{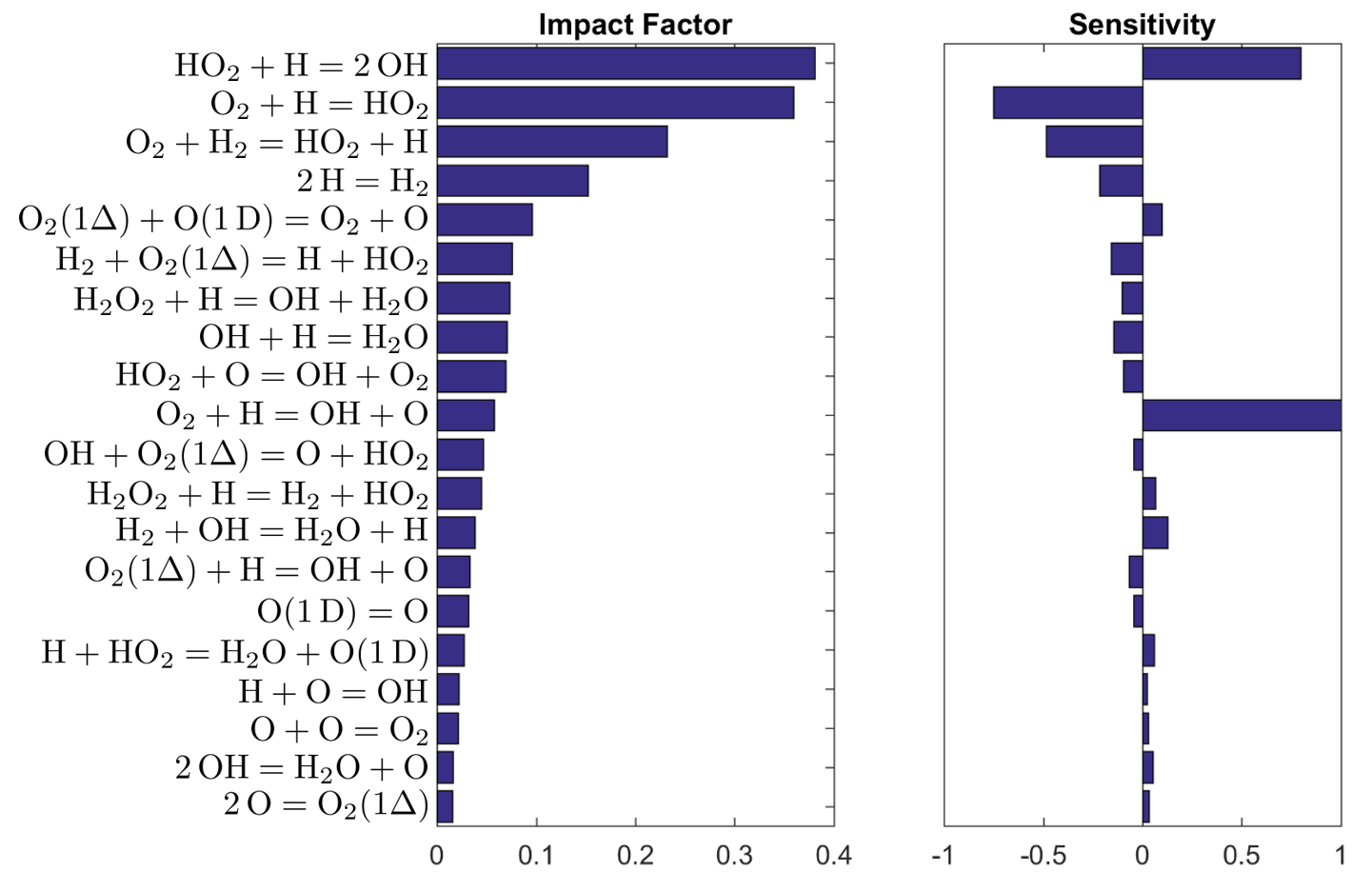}
	\caption{Sensitivity and ranked impact-factor results for a laminar flame speed QOI (a00000484).
		The top 20 reactions are ranked by impact factor, defined as absolute sensitivity multiplied by
		0.5log$(U_e/L_e)$.}
	\label{fig:impact}
\end{figure}

The workflow approach can reveal chemistry or kinetics not being captured, potentially important for combustion technology but also providing a new mechanistic insight. Such appears to be the case for the role of \mbox{$\text{O}_2(\text{a}^1\Delta\text{g})$} here. Comparisons of Mechanism3 vs. Mechanism2 revealed significant differences ($\sim$15\%) in flame-speed predictions for high-pressure, fuel-rich flames due to the added electronic-excited-state chemistry from Konnov \cite{konnov2015role}. Further examination, prompted by \mbox{\ce{H2 + }\mbox{\ce{O2}$(1\Delta)$}\ce{= H + HO2} } having among the highest sensitivity coefficients (Figure \ref{fig:impact}), suggested that the reverse reaction \mbox{\ce{H + HO2 = H2 + }\mbox{\ce{O2}$(1\Delta)$} } serves as a pathway for \ce{H2 + O2} production, supplementing the direct reaction to ground-state products, \ce{H + HO2 = H2 + O2($^3$P)}. The latter reaction is known to act as a significantly inhibitive chain-terminating reaction in high-pressure flames \cite{burke2010negative}, consistent with the explanation of a 15\% lower flame speed for the Mechanism3 compared to the Mechanism2.

Additional simulations used a modified version of the Mechanism2 where the \ce{H + HO2 = H2 + O2} reaction rate constant was scaled to include the additional flux through this \mbox{$\text{O}_2(\text{a}^1\Delta\text{g})$} pathway, implicitly assuming that quenching to the ground state was fast. These simulations yielded identical predictions (within 1\%) compared to Mechanism3, which had included the full electronic-excited-state chemistry set.

Interestingly, while predictions using model parameters at their nominal values suggested minimal contributions from other (non-quenching) \mbox{$\text{O}_2(\text{a}^1\Delta\text{g})$} pathways, the uncertainty analysis (Figure \ref{fig:impact}) indicated that these other \mbox{$\text{O}_2(\text{a}^1\Delta\text{g})$} pathways, including \mbox{\ce{H + O2}$(1\Delta)$\ce{ = OH + O}}, are indeed among the most significant contributors to prediction uncertainties and, consequently, to experimental interpretations. More subtly, the present treatment converts a structural uncertainty to a parametric uncertainty (similar to \cite{russi2008sensitivity}). In this case, a structural uncertainty regarding the implicit assumption of a particular electronic state formed from \ce{H + HO2} or of rapid quenching of the nascent \mbox{$\text{O}_2(\text{a}^1\Delta\text{g})$} is converted into parametric uncertainties in rate coefficients of the individual reactions within the overall kinetic sequence.

Achieving more definitive conclusions regarding the role of \mbox{\ce{H2 +}\mbox{\ce{O2}$(1\Delta)$}\ce{= H + HO2}} and subsequent \mbox{$\text{O}_2(\text{a}^1\Delta\text{g})$} consumption pathways would benefit from reinterpretations of the raw data from the studies used to derive \ce{H + HO2} rate constants and product branching ratios \cite{sridharan1982kinetics, keyser1986absolute, baldwin1979rate}, perhaps within theory-based, multi-scale informatics frameworks \cite{burke2013quantitative, burke2015multiscale} to ensure consistent, physically rigorous interpretations of limited experimental measurements for these multi-channel reactions. Nevertheless, the present analyses suggest an unexpected role of \mbox{$\text{O}_2(\text{a}^1\Delta\text{g})$} chemistry in high-pressure flames.

\subsection{Detecting data inconsistencies}
Consistency analysis of the entire dataset revealed that several experimental low-temperature ($< 1000$ K), high-pressure ($\sim 3-4$ atm) shock-tube ignition-delay times were individually inconsistent within the model structure and kinetic-model uncertainty assignments (without considering other experimental targets). For those QOIs, predictions of the \ce{OH} maximum (used as the ignition target) were at least 2.5 times longer than measured experimentally. Shock-tube ignition delay times at these conditions are known to require more sophisticated physical models that consider heterogeneous gas-dynamic and heat-release effects \cite{meyer1971shock, pang2009experimental, ihme2015detailed}, going beyond the simple, homogeneous, constant-volume reactor model employed here. Proper consideration of such QOIs within UQ frameworks requires the use of these more-appropriate physical models and any additional accompanying uncertainties in the initial and boundary conditions for these experiments.

This finding shows that a broader impact of this approach is to provide a means of uncovering which experiments and parameters are inconsistent with the larger body of work from the community and thus of designing new experiments and parameter calculations. Such a finding could mean merely that the experiment does not match the idealized physical model being used. Rather than being discarded as ``bad data,'' the data can become valuable through identifying the need for more information to be collected or for a different physical model.

\subsection{Role of a workflow for dynamic mechanism development}
Comparing experiments with each other or comparing model predictions to data is complicated by the amount of data, the oft-times complex models, and various uncertainties. Stochastic (aleatoric) uncertainties are present in raw data, and systematic, nonstatistical (epistemic) uncertainties are present in the analyzed data due to uncertainties in calibrations, data-analysis procedures, apparatus-to-apparatus variations, and imperfect understanding of the experimental system's transport, chemistry, and physical-environment influences.

For a single quantity, such as a rate coefficient or enthalpy of formation, values may be compared using their significant digits and/or with a Gaussian-distribution’s sample standard deviation $\sigma$. Two-dimensional graphs allow comparisons of data and/or predictions (with their uncertainties) for a single independent variable, and multiple graphs allow multiple comparisons. If a second independent variable is added, a three-dimensional surface can be displayed, but a comparison of one surface to another is difficult and uncertainty becomes quite difficult to display.

The workflow described here, applied to a simple combustion system that has both importance and unresolved discrepancies among data, provides a systematic and mathematically sound approach to assessing the agreement of models with uncertainty vs. experiments with uncertainty.

Identification of the reactions and experiments central to achieving model-data consistency and quantifying uncertainties based on present knowledge thereafter guides the portions of the UQ framework to be further refined. Further refinement of the uncertainty assignments for important data is essential to the UQ analysis. The inclusion of temperature-dependent, pressure-dependent, and bath-gas-dependent rate constant uncertainties \cite{turanyi2012determination, burke2013quantitative, sheen2013kinetics} is essential for accurate rate-parameter uncertainty quantification. The inclusion of elementary kinetics theories \cite{burke2013quantitative, burke2015multiscale} is essential for physically rigorous interpretations and reliable extrapolations based on limited experimental data. The inclusion of transport parameters is essential to reveal sensitivity to them and to different transport models. Finally, inclusion of physical/instrumental model uncertainties \cite{burke2013quantitative, yeates2015integrated} is essential to quantify the information content of experimental data.

\section{Conclusion}
Analysis of large combustion-data sets with data-sciences tools was applied for evaluation and improvement of \ce{H2}/\ce{O2} combustion, a foundation of combustion modeling. A cloud-based workflow aided comparison of models to shock-tube and flame data, incorporating the uncertainties in each. Thermochemistry and kinetics parameters were more easily compared across the literature, and the inhibiting reaction \mbox{\ce{H + HO2 = H2 +  O2}$(1\Delta)$} was identified as a pathway for \ce{H2 + O2($^3$P)} production in high-pressure flames, supplementing the direct reaction to ground-state products. In addition, a dynamically improvable new chemical model was developed for \ce{H2}/\ce{O2} combustion, optimized within acceptable model-parameter uncertainties.

Finally, the present analyses contribute to the growing set of examples that UQ can be a powerful tool for scientific investigations into model parameters and structures. The collaborative-workflow approach represents an update of the GRI-Mech approach that can engage a larger breadth of the combustion-kinetics community, as well as more complex data sets and models.

\section{Acknowledgments}
We gratefully acknowledge financial support by U.S. Department of Energy, National Nuclear Security Administration, under Award Number $\text{DE}$-$\text{NA}$0002375 (JO/MF); U.S. National Science Foundation award 1340609/Supplement (CDN/PRW) and grant number 1403171 (RHW); Saudi Aramco under the FUELCOM program and by competitive research funding from King Abdullah University of Science and Technology (SB/SMS); Columbia University (MPB); and the U.S. Air Force Office of Scientific Research for past support and the Deutsches Zentrum f{\"u}r Luft- und Raumfahrt for current support of the PrIMe Data Warehouse.

\newpage
\raggedright

\interlinepenalty=10000
\bibliography{library}


\end{document}